\documentclass[aps,prl,twocolumn,groupedaddress]{revtex4-1}
\usepackage[]{latexsym,amsmath,amssymb,graphicx,epsfig}
 \usepackage{longtable}
\newcommand{\be}{\begin{eqnarray}}
\newcommand{\ee}{\end{eqnarray}}

\begin{document}

%\preprint{}

\title{Observable Proton Decay from Planck Scale Physics}


\author{S.M. Barr}
\email[]{smbarr@bartol.udel.edu}
\affiliation{Department of Physics and Astronomy, Bartol Research Institute, University of Delaware, Newark, Delaware 19716, USA}
\author{Xavier Calmet}
\email[]{x.calmet@sussex.ac.uk}
%\thanks{}
\affiliation{Department of Physics $\&$ Astronomy, University of Sussex, Falmer, Brighton, BN1 9QH, UK}


\date{\today}

\begin{abstract}
In the Standard Model, no dim-5 $\Delta B \neq 0$ operators exist,
so that Planck-scale-induced proton decay amplitudes are suppressed
by at least $1/M^2_{P \ell}$. If the Standard Model is augmented by
a light, color-non-singlet boson, then $O(1/M_{P \ell})$
proton-decay amplitudes are possible. These always conserve $B+L$,
so that the dominant decay modes are $p \rightarrow \Pi^+ \nu$ and
$p \rightarrow \Pi^+ \Pi^+ \ell^-$, where $\Pi^+ = \pi^+$ or $K^+$.
\end{abstract}

\pacs{}
%\keywords{}

\maketitle

\section{Introduction}
One of the most serious obstacles confronting particle physics is
the extreme practical difficulty of probing the physics of the
Planck scale. One way Planck-scale physics might be probed is by
means of proton decay \cite{Harnik:2004yp}. There are general
arguments that quantum gravity should violate global quantum
numbers, such as baryon number ($B$) and lepton number ($L$).
Therefore, proton decay should happen as a consequence of
Planck-scale physics. Of course, proton decay could also happen as a
result of grand unification, and grand unification is an extremely
well-motivated idea. Nevertheless, it is not absolutely certain that
the idea of grand unification is correct, and even if it is, there
could be other contributions to proton decay. Therefore, if proton
decay is observed, we should keep an open mind about what is causing
it. As we shall see below, if it is coming from Planck-scale physics
certain kinds of colored bosons should exist whose mass should be
less than $10^7$ GeV and very likely much less and within reach of
accelerators, and certain decay modes of the proton should be seen.
Moreover, there would be a relation between the kind of light
colored bosons and the proton decay modes.

At first glance, it would seem hopeless to see proton decay caused by Planck-scale
physics. If the operators violating baryon number were dimension-6, with a
suppression factor of order $M_{P \ell}^{-2}$, the proton-decay rate
would be much too small to be seen in the foreseeable future, or perhaps ever.
On the other hand, if the operators were dimension-4, proton decay would
typically be too fast. So it is dimension-5 $B$-violating operators
suppressed by a single power of $M_{P\ell}$ that are of interest in this regard.
No such operators can be constructed from Standard Model (SM) fields
alone, but they can be if non-Standard
Model fields exist. One possibility, already explored in \cite{Harnik:2004yp},
is that such operators may arise in the context of low-energy supersymmetry.
In this paper we shall show that even
in much more modest extensions of the Standard Model proton-decay
amplitudes can arise
at $O(1/M_{P \ell})$. In particular, we shall show that adding just a
single type of non-SM field (specifically, colored bosons, as noted above)
allows this possibility.
We shall classify all such cases and the resulting effective
$B$-violating operators, somewhat in the spirit of the
classic analysis of Weinberg and of Wilczek and Zee
\cite{Weinberg:1979sa,Wilczek:1979hc}.

Let us call the single new field added to the Standard Model $X$. If
$X$ has both dim-5 $B$-violating couplings and also renormalizable
couplings to SM fields, then integrating it out will yield effective
$B$-violating operators involving only SM fields that are suppressed
by $M_{P\ell}^{-1} M_X^{-2}$, if $X$ is a boson. Thus, as we will
show later, for there to be observable proton decay, $M_X$ would
have to be less than about $10^7$ GeV, and probably even lighter,
since there are likely to be small dimensionless couplings involved.
We shall call such fields ``light" in what follows.

A dim-5 operator would have to be either (i) a product of five boson
fields or derivatives, or (ii) a fermion bilinear times a product of
two boson fields or derivatives. Suppose $X$ were a fermion field.
Then the only light boson would be the SM Higgs doublet $\Phi$, and
therefore no operator of type (i) would exist. And for type (ii),
the fermion bilinear would have to be a color-singlet, so that by
assigning $B = \frac{1}{3}$ to all color-triplet fermions and $B = -
\frac{1}{3}$ and to all color anti-triplet fermions $B$ would be
conserved.

Consequently, in order to violate $B$ at dim-5, $X$ has to be a
boson field. Moreover, it is not difficult to see from reasoning
similar to the above that it must be a color non-singlet. There are
only ten color non-singlet scalars that can couple renormalizability to
the fermions of the Standard Model. The possibilities are listed in
Table I.

In the second column of Table I, we show all the dim-4 couplings of
$X$ and $X^{\dag}$ to the SM fermions. The left-handed fermions of
the Standard Model are denoted in Table I and throughout this paper
by $Q$, $L$, $u^c$, $d^c$, and $\ell^+$. In certain cases, not all
the dim-4 Yukawa couplings permitted by the gauge symmetries of the
Standard Model can be present in the Lagrangian without allowing
catastrophic proton decay (i.e. proton decay that is not suppressed
by $1/(M_{P \ell} M_X^2)$, but only by $1/M_X^2$). These cases are
therefore divided in Table I into subcases A and B (denoted by a
subscript in column 1), depending on which dim-4 operators are
present. Those not present must be forbidden by some symmetry beyond
the SM gauge group.

In column 3 of Table I, we show those dim-5 couplings of $X$ and
$X^{\dag}$ to the SM fermions that violate baryon number and are
linear in $X$ or $X^{\dag}$. (An operator quadratic in $X$
would lead to a proton-decay rate of order $\left(
\frac{m_p^{5}}{M_{P \ell} M_X^4} \right)^2 m_p$ because of the need
to integrate out two $X$ bosons instead of one.) In the last column
of Table I, we show the dim-7 four-fermion, $\Delta B \neq 0$ operators
that arise from integrating out the $X$ field. A general analysis of
such dim-7 operators was carried out in \cite{w-wz-1980}.
In the operators of
Table I, the fermions in parentheses are contracted into Lorentz
scalars; the dots represent a contraction of $SU(2)_L$ doublets into
singlets; and the circles represent the contraction of $SU(2)_L$
doublets into triplets.

A noteworthy feature of all the cases is that $B+L$ is conserved
rather than $B-L$. ($B+L$ conserving baryon decay was analyzed in
the context of $R$-parity violating supersymmetric models by Vissani
in \cite{Vissani} and more recently by Babu and Mohapatra in a large
class of GUT models \cite{Babu:2012vb}.) Consequently, the dominant
proton decay decay modes are $p \rightarrow \Pi^+ \nu$, $p
\rightarrow \Pi^+ \Pi^+ \ell^-$, where $\Pi^+ = \pi^+$ or $K^+$. Of
course, in practice one cannot tell in the difference between the $p
\rightarrow \Pi^+ \nu$ (or $n \rightarrow \Pi^0 \nu$), which occur
in these models, and $p \rightarrow \Pi^+ \overline{\nu}$ (or $n
\rightarrow \Pi^0 \overline{\nu}$), which arise in typical grand
unified theories(GUTs). However, in GUTs these are accompanied by
decay modes with positively charged anti-leptons, whereas in models
we are looking at they are not. In Table I, the operators that cause
$p \rightarrow \Pi^+ \nu$ are called type (a), while those that
cause $p \rightarrow \Pi^+ \Pi^+ \ell^-$ are called type (b). For
neutron decay the dominant modes are $n \rightarrow \Pi^0 \nu$ for
type (a), and $n \rightarrow \Pi^+ \ell^-$ for type (b).

%\vspace{0.2cm}

%\vspace{0.2cm}

\begin{table*}
\begin{tabular}{|l|l|l|l|}
\hline {\bf SM rep of $X$} & dim-4  & $\Delta B \neq 0$, dim-5 & effective 4-fermion operators \\
\hline $(3,2, \frac{1}{6})$ & $X (d^c \;L)$ & (a) $\Phi \cdot X (Q
\cdot Q)$ & $O_1 = (Q \cdot Q)^{\dag} (d^c L \cdot \langle
\Phi^{\dag}
\rangle) = (u\; d)^{\dag} (d^c \nu)$ \\
" & " & (a,b) $X \cdot (Q \;Q) \cdot \Phi$ & $O'_1 =  \langle
\Phi^{\dag} \rangle \cdot (Q^{\dag} Q^{\dag})
\cdot (L d^c)$ \\
& & & $\;\;\;\;\;\; = (u \; d)^{\dag} (\nu d^c), (d \;
d)^{\dag}(\ell^- d^c)$ \\
\hline $(\overline{3}, 1, - \frac{2}{3})$ & $X (d^c \; d^c)$ & (b)
$X \Phi \cdot (Q \; \ell^+)$
& $O_2 = \langle \Phi^{\dag} \rangle \cdot (Q \; \ell^+)^{\dag} (d^c \; d^c) = (d \ell^+)^{\dag} (d^c \; d^c) $  \\
" & " & (b) $X ^{\dag} \Phi \cdot (L \; d^c)$ & $O_3 = \langle \Phi
\rangle \cdot (L \; d^c)(d^c \; d^c)
= (\ell^- d^c) (d^c \; d^c)$ \\
" & " &  (a) $X^{\dag} \Phi^{\dag} \cdot (L \; u^c)$ & $O_4 =
\langle \Phi^{\dag} \rangle \cdot (L \; u^c)(d^c \; d^c)
= (\nu u^c) (d^c \; d^c)$ \\
\hline $( \overline{3}, 1, \frac{1}{3})_A$ & $X(Q \; L)$, $X^{\dag} (u^c \; \ell^+)$ & --- & --- \\
\hline $( \overline{3}, 1, \frac{1}{3})_B$ & $X(u^c \; d^c)$ & (a)
$X^{\dag} \Phi^{\dag} \cdot (L \; d^c)$ & $O_5 = \langle \Phi^{\dag}
\rangle \cdot (L \; d^c)(u^c \; d^c)
= (\nu d^c) (u^c \; d^c)$ \\
" & $X^{\dag}(Q \cdot Q)$ & (a) $X^{\dag} \Phi^{\dag} \cdot (L \;
d^c)$
& $O_1$ \\
\hline $( 3, 3, -\frac{1}{3})_A$ & $X^{\dag}(Q \circ L)$ & --- & --- \\
\hline $( 3, 3, -\frac{1}{3})_B$ & $X(Q \circ Q)$ & (a,b) $X
\Phi^{\dag} \circ (L \; d^c)$ & $O^{''}_1 = (Q \circ Q)^{\dag} (d^c
L) \circ \langle \Phi^{\dag}
\rangle$ \\
& & & $\;\;\;\;\;\; = (u\; d)^{\dag} (d^c \nu), (d\; d)^{\dag}(d^c \ell^-)$\\
\hline $( \overline{6}, 1, -\frac{1}{3})$ & $X(Q \cdot Q)$ & --- & --- \\
\hline $( \overline{6}, 3, -\frac{1}{3})$ & $X(Q \circ Q)$ & --- & --- \\
\hline $( 3, 2, \frac{7}{6})$ & (a) $X \cdot (L \; u^c)$ &
$\Phi^{\dag} \cdot
X^{\dag} (d^c \; d^c)$ & $O_4$ \\
" & (b) $X^{\dag}\cdot (Q \; \ell^+)$ & $\Phi^{\dag} \cdot X^{\dag}
(d^c
\; d^c)$ & $O_2$ \\
\hline $( 3, 1, -\frac{4}{3})_A$ & $X(d^c \; \ell^+)$ & --- & --- \\
\hline $( 3, 1, -\frac{4}{3})_B$ & $X^{\dag} (u^c \; u^c)$ & --- & --- \\
\hline $( \overline{6}, 1, -\frac{4}{3})$ & $X^{\dag}(u^c \; u^c)$ & --- &  --- \\
\hline $( \overline{6}, 1, \frac{2}{3})$ & $X^{\dag} (d^c \; d^c)$ & --- & --- \\
\hline
\end{tabular}
\caption{The possibilities for
color-non-singlet scalars ($X$), their dim-4 and $\Delta B \neq 0$
dim-5 couplings to the Standard Model fermions, and the $\Delta B
\neq 0$ four-fermion operators that arise from integrating out the
$X$. The operators that cause $p \rightarrow \Pi^+ \nu$ are called
type (a), while those that cause $p \rightarrow \Pi^+ \Pi^+ \ell^-$
are called type (b).}
\end{table*}
%\vspace{0.5cm}

Let us consider, for example, the first case in Table I, where $X$
is a boson with the same SM quantum numbers as the left-handed quark
doublet, that is, a $(3,2, \frac{1}{6})$ of $SU(3)_c \times SU(2)_L
\times U(1)_Y$. In this case, there is only one dim-4 coupling to
the SM quarks and leptons, namely $X(d^c \; L)$, which requires that
$X$ have $B= \frac{1}{3}$ if $B$ is to be conserved by
renormalizable couplings. On the other hand, there are several dim-5
operators involving $X$. Of these, $\Phi \cdot X (u^c \; \ell^+)$
conserves $B$; $X^{\dag} \cdot X^{\dag} (u^c \; \ell^+)$ violates
$B$ but is quadratic in the $X$ field; and operators of the form
$X^{\dag} X^{\dag} (Q \; Q)$ are both quadratic in $X$ and conserve
$B$. That leaves only the dim-5 operators $\Phi \cdot X (Q \cdot Q)$
and $X \cdot (Q \;Q) \cdot \Phi$ shown in the first and second rows
of Table I that can lead to observable proton decay. When the $X$
field is integrated out, through the diagram shown in Fig. 1, one
obtains the $B$-violating four-fermion operators called $O_1$ and
$O'_1$ in Table I. These exotic scalar fields have been studied
extensively, see e.g. \cite{Ma:1998pi,Gu:2011pf,Vecchi:2011ab,Giudice:2011ak}.
In particular, in it is pointed out in \cite{Vecchi:2011ab,Giudice:2011ak} that
they could have interesting consequences for physics at the CERN Large Hadron Collider if their masses are low enough.

\begin{figure}[ht]
\begin{center}
\includegraphics[width=5cm]{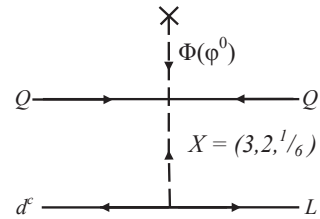}
\end{center}
\caption{Integrating $X$ gives a $\Delta B = - \Delta L$ operator
with coefficient of $O(\langle \Phi \rangle/(M_{P \ell} M_X))$.}
\end{figure}

%\vspace{0.2cm}

If we examine the four-fermion operators in the last column of Table
I, we see that all are of the general form $\overline{d}
\overline{d} \overline{d} \ell^-$ or $\overline{u} \overline{d}
\overline{d} \nu$, where $\overline{d}$ and $\overline{u}$ stand for
anti-quarks without respect to handedness. These can be dressed by
sea quarks to give proton decay as shown in Fig. 2.

\begin{figure}[ht]
\begin{center}
\includegraphics[width=9cm]{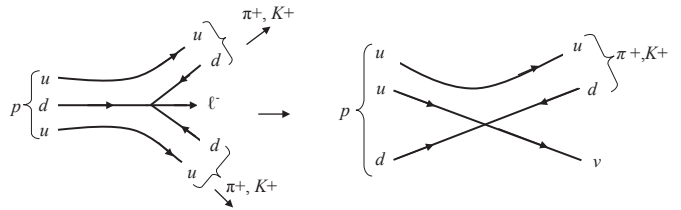}
\end{center}
\caption{The four-fermion operators in the last column of Table I can
be dressed by sea quarks to give proton decay.}
\end{figure}

From Table I, we see that there are five kinds of models, defined by
the quantum numbers of $X$ and its couplings: Model 1: $X = (3,2,
\frac{1}{6})$; Model 2: $X = (\overline{3}, 1, - \frac{2}{3})$;
Model 3: $X = (\overline{3}, 1, - \frac{1}{3})_B$, Model 4: $X =
(\overline{3}, 3, - \frac{1}{3})_B$; and Model 5: $X = (3,2,
\frac{7}{6})$. In Model 1, all the decay modes shown in Fig. 2 are
allowed. However, in Model 2 (and Model 5, which has the same
four-fermion proton decay operators), the anti-down quarks in the
factor $(d^c \; d^c)$ of the operators $O_2$ and $O_4$ must be
antisymmetric in flavor due to Fermi statistics (since they are
antisymmetric in both spin and color) and thus contain a strange
quark. Thus the dominant decays are $p \rightarrow K^+ \nu$ and $p
\rightarrow K^+ \pi^+ \ell^-$, while the modes with pions and no
Kaons must come from higher-order effects. (The operator $O_3$ must
have all three of its anti-down quarks in a flavor antisymmetric
state, and so must contain $d^c s^c b^c$, which means that it does
not contribute to proton decay, except through higher-order
effects.)

Model 3 gives the decay modes $p \rightarrow \pi^+ \nu$ and $p
\rightarrow K^+ \nu$, but does not give decay modes with charged
leptons except through higher-order effects. Model 4 gives the decay
modes $p \rightarrow \pi^+ \nu$, $p \rightarrow K^+ \nu$, and $p
\rightarrow K^+ \pi^+ \ell^-$, but the mode $p \rightarrow \pi^+
\pi^+ \ell^-$ would only arise at higher-order, as the product $(d\;
d)^{\dag}$ appearing in the operator $O^{''}_1$ is flavor
antisymmetric. The dominant nucleon decay modes in the five models
are shown in Table II, while in Table III are listed the present
experimental limits for those modes \cite{pdg}. Note that these
limits all lie in the range (0.17 to 6.7) $\times 10^{32}$ yrs.

\begin{table*}
\begin{tabular}{|l|l|l|}
\hline {\bf SM rep of $X$} & dominant $p$ decay modes & dominant $n$ decay modes \\
\hline $(3,2, \frac{1}{6})$ & $p \rightarrow \pi^+ \nu$, $K^+ \nu$,
$\pi^+ \pi^+ \ell^-$, $K^+ \pi^+ \ell^-$ & $n \rightarrow \pi^0
\nu$, $K^0 \nu$,
$\pi^+ \ell^-$, $K^+ \ell^-$, $K^0 \pi^+ \ell^-$\\
\hline $(\overline{3}, 1, - \frac{2}{3})$ & $p \rightarrow K^+ \nu$,
$K^+ \pi^+ \ell^-$ & $n \rightarrow K^0 \nu$, $K^+ \ell^-$, $K^0 \pi^+ \ell^-$ \\
\hline $( \overline{3}, 1, \frac{1}{3})_B$ & $p \rightarrow \pi^+
\nu$, $K^+ \nu$ & $n \rightarrow \pi^0 \nu$, $K^0 \nu$ \\
\hline $( 3, 3, -\frac{1}{3})_B$ & $p \rightarrow \pi^+ \nu$, $K^+
\nu$, $K^+ \pi^+ \ell^- $ & $n \rightarrow \pi^0 \nu$,
$K^0 \nu$, $K^+ \ell^-$,
$K^0 \pi^+ \ell^-$ \\
\hline $( 3, 2, \frac{7}{6})$ & $p \rightarrow K^+ \nu$, $K^+ \pi^+
\ell^-$ & $n \rightarrow K^0 \nu$,
$K^0 \pi^+ \ell^-$, $K^+ \ell^-$ \\
\hline
\end{tabular}
\caption{The dominant nucleon decay modes in the five models.}
\end{table*}

\begin{table*}
\begin{tabular}{|l|l|}
\hline {\bf nucleon decay mode} & experimental limits  \\
\hline $p \rightarrow \pi^+ \nu$ & $0.25 \times 10^{32}$ yrs \\
\hline $p \rightarrow K^+ \nu $ & $6.7 \times 10^{32}$ yrs \\
\hline $p \rightarrow \pi^+ \pi^+ e^-$ & $0.3 \times 10^{32}$ yrs \\
\hline $p \rightarrow \pi^+ \pi^+ \mu^-$ & $0.17 \times 10^{32}$ yrs \\
\hline $p \rightarrow K^+ \pi^+ e^-$ & $0.75 \times 10^{32}$ yrs \\
\hline $p \rightarrow K^+ \pi^+ \mu^-$ & $2.45 \times 10^{32}$ yrs
\\
\hline $n \rightarrow \pi^0 \nu$ & $1.12 \times 10^{32}$ yrs \\
\hline $n \rightarrow K^0 \nu$ & $0.86 \times 10^{32}$ yrs \\
\hline $n \rightarrow \pi^+ e^-$ & $0.65 \times 10^{32}$ yrs
\\
\hline $n \rightarrow \pi^+ \mu^-$ & $0.49 \times 10^{32}$ yrs
\\
\hline $n \rightarrow K^+ e^-$ & $0.32 \times 10^{32}$ yrs
\\
\hline $n \rightarrow K^+ \mu^-$ & $0.57 \times 10^{32}$ yrs
\\
\hline $n \rightarrow \pi K \ell^-$ & - \\
\hline
\end{tabular} \caption{The present 90 percent confidence
limits on the nucleon decay modes
shown in Table II.}
\end{table*}

Suppose that we denote by $(Y_4)_{ab}$ and $(Y_5)_{ab}/M_{P\ell}$
the coefficients of the dim-4 and dim-5 operators shown in the
second and third columns of Table I. These are matrices in flavor
space, and the indices $a,b$ denote the families of the fermions in
these operators. Since we don't know the flavor dependence of these
matrices, for the purpose of making rough estimates let us denote
the typical value of these couplings to the fermions of the first
and second families simply by $Y_4$ and $Y_5/M_{P\ell}$, without
indices. Then the effective four-fermion $\Delta B = - \Delta L$
operators in the last column of Table I have coefficients of the
form $Y_4 Y_5 \frac{v}{M^2_X M_{P \ell}}$, where $v$ is the vacuum
expectation value of the Standard Model Higgs field $\Phi$. A
two-body nucleon decay mode produced by such operators would
therefore typically have a partial rate of order

\begin{eqnarray}
\Gamma &\sim& \frac{1}{16 \pi} \left[ Y_4 Y_5
\frac{v/\sqrt{2}}{M_X^2 M_{P \ell}} \right]^2 m_p^5 \\ \nonumber
&\sim& (Y_4 Y_5)^2 \left( \frac{M_X}{10^7 {\rm GeV}} \right)^{-4} (3
\times 10^{-32} {\rm yr}^{-1}).
\end{eqnarray}

\noindent Thus, to produce two-body nucleon-decay partial rates near
the present limits (which, as noted, are all $[k \times 10^{32} {\rm
yr}]^{-1}$ (with $0.17 \leq k \leq 6.7$) would require

\begin{equation}
M_X \sim (3k)^{1/4} (Y_4 Y_5)^{1/2} 10^7 {\rm GeV}.
\end{equation}

\noindent By analogy with the Yukawa couplings of the Standard Model
Higgs field, it is plausible that the coefficients $Y_4$ and $Y_5$
to the first and second families could be of order $10^{-5}$ to
$10^{-3}$, so that the $X$ bosons in these models could be in range
of detection at accelerators. Such light, colored bosons raise the
question of the whether they would be in conflict with bounds on
flavor-changing neutral current (FCNC) processes such as $K
\overline{K}$ mixing and $\mu \rightarrow e \gamma$.

The contribution of the $X$ scalars through box diagrams to the
coefficients of $\Delta S = 2$ operators would be of order
$\frac{1}{16 \pi^2} (Y_4)^4/M_X^2$. To avoid excessive CP violation
in $K \overline{K}$ system, this must be less than about $10^{-15}
{\rm GeV}^{-2}$. This gives $M_X > (Y_4)^2 (2.5 \times 10^6$ GeV).
For models 1 and 5, where $X$ is a leptoquark, one-loop diagrams
involving virtual $X$ bosons would typically contribute to $\mu
\rightarrow e \gamma$. The present limits for this process would
require $M_X > Y_4 (10^5$ GeV). More stringent limits on these
leptoquark masses come from $K_L \rightarrow \mu^{\pm} e^{\mp}$.
Present limits for this would give $M_X > Y_4 (2 \times 10^6$ GeV).
Of course, these are rough estimates. Special patterns of flavor
dependence of the matrices $(Y_4)_{ab}$ and $(Y_5)_{ab}$ could
either relax or strengthen these limits significantly.

It is apparent that these models could allow observable nucleon
decay rates, while satisfying limits from FCNC processes. For
instance, if $M_X \sim 10$ TeV, the FCNC limits would be satisfied
if $Y_4 < 5 \times 10^{-3}$, while observable two-body nucleon decay
could occur if $Y_4 Y_5 \sim 10^{-6}$. Similarly, if $M_X \sim 1$
TeV, then the FCNC limits would require $Y_4 < 5 \times 10^{-4}$,
while observable two-body proton decay could occur for $Y_4 Y_5 \sim
10^{-8}$.

Because one does not know {\it a priori} the flavor dependence of
the coefficients $Y_4$ and $Y_5$, one cannot say for nucleon decay
whether pion or kaon modes will be dominant, and whether electron or
muon modes will be. All else being equal, phase space would lead one
to expect that for proton decay the decay modes with neutrinos,
which are two-body, would dominate over the decay modes with
charged-leptons, which are three-body.
However, in models 2 and 5, different operators produce the neutrino
and charged lepton modes. For example, in model 2 the neutrino modes
come from the dim-5 operator $X^{\dag} \Phi^{\dag} \cdot L u^c$,
whereas the charged lepton modes come from the dim-5 operator $X
\Phi \cdot Q \ell^+$. And one does not know {\it a priori} which of
these operators has larger coefficients. In models 1 and 4, on the
other hand, the operators $O'_1$ and $O^{''}_1$, contribute to both
neutrino and charged lepton decay modes of the proton.
So the former should predominate
because of phase space. In model 3, only neutrino modes are produced
by the operators $O_1$ and $O_5$.

It is interesting to compare the kinds of non-supersymmetric
models discussed here with
models based on low-energy supersymmetry that can also give
proton-decay amplitudes of order $1/M_{P\ell}$, as discussed
in \cite{Harnik:2004yp}. There are several differences. First, one notes
that in \cite{Harnik:2004yp}
the baryon-number-violating operators are constructed out of
chiral superfields that appear in the Minimal Supersymmetric Standard
Model (MSSM). While the first three cases we consider in Table II
($(3, 2, \frac{1}{6})$, $(\overline{3}, 1, - \frac{2}{3})$, and
$({\overline3}, 1, \frac{1}{3})$) correspond to fields that exist in
the MSSM, the other two cases do not. Second, one sees that
even in the cases where our $X$
field corresponds to a field of the MSSM, many of the dim-5
$\Delta B \neq 0$ operators in Table I, and in particular those containing
$X^{\dag}$, have no analogue in the MSSM. For example, in the case
$X = (\overline{3}, 1, -\frac{2}{3})$, there are the operators
$X^{\dag} \Phi \cdot (L \; d^c)$ and $X^{\dag} \Phi^{\dag} \cdot ( L \;
u^c)$. As far as gauge quantum numbers go, these would correspond
to MSSM operators of the form $u^{c \dag} H_d L d^c$ and
$u^{c \dag} H_u L u^c$. Neither of these is an $F$ term, however, and
therefore they cannot appear in an effective superpotential.
The same is true of
the operators $O_1$, $O'_1$, and $O^{''}_1$ in the last column of Table I.
The most dramatic difference is that in the models discussed here,
proton decay conserves $B+L$ rather than $B-L$ as in the supersymmetric
models of \cite{Harnik:2004yp}. This would not give a discernible
difference for neutrino modes, since one cannot tell the difference
in practice between, for example, $p \rightarrow K^+ \overline{\nu}$
(which arises in the MSSM) and $p \rightarrow K^+ \nu$ (which arises
in the present models). In the MSSM, however, the anti-neutrino modes
would be accompanied by modes with positively charged anti-leptons,
which would not appear in the Planck-scale proton decay models discussed
here.

As we are examining these questions ``from the bottom up", we do not
attempt to explain why the colored bosons we have called $X$ should
be light compared to the Planck scale. It is, however, interesting
that if such bosons are lighter than about $10^7$ GeV, they may
allow a direct window onto physics at the Planck scale.

\begin{acknowledgments}
S.M.B. acknowledges the support of Department of Energy grant
DE-FG02-91ER40626.
\end{acknowledgments}


\end{document}